\documentclass[twoside,twocolumn,9pt]{article}
\usepackage{extsizes}
\usepackage[super,sort&compress,comma]{natbib}
\usepackage[version=3]{mhchem}
\usepackage[left=1.5cm, right=1.5cm, top=1.785cm, bottom=2.0cm]{geometry}
\usepackage{balance}
\usepackage{times,mathptmx}
\usepackage{sectsty}
\usepackage{graphicx}
\usepackage{lastpage}
\usepackage[format=plain,justification=justified,singlelinecheck=false,font={stretch=1.125,small,sf},labelfont=bf,labelsep=space]{caption}
\usepackage{float}
\usepackage{fancyhdr}
\usepackage{fnpos}
\usepackage[english]{babel}
\usepackage{array}
\usepackage{charter}
\usepackage[T1]{fontenc}
\usepackage[usenames,dvipsnames]{xcolor}
\usepackage{setspace}
\usepackage[compact]{titlesec}
\usepackage{hyperref}

\usepackage{epstopdf}

\definecolor{cream}{RGB}{222,217,201}

\begin{document}

\pagestyle{fancy}
\thispagestyle{plain}
\fancypagestyle{plain}{

\fancyhead[C]{\includegraphics[width=18.5cm]{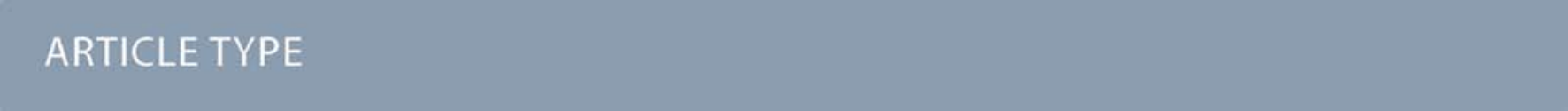}}
\fancyhead[L]{\hspace{0cm}\vspace{1.5cm}\includegraphics[height=30pt]{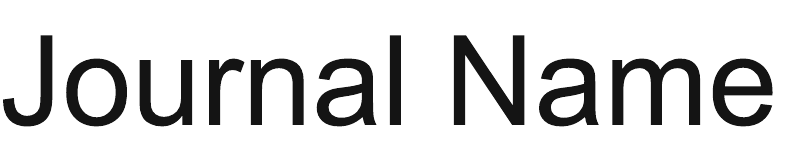}}
\fancyhead[R]{\hspace{0cm}\vspace{1.7cm}\includegraphics[height=55pt]{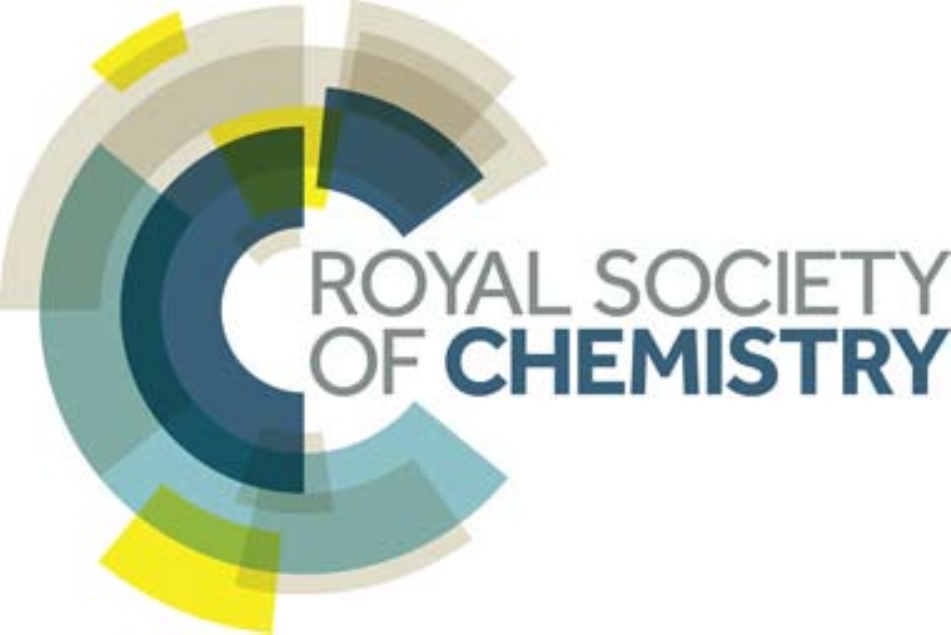}}
\renewcommand{\headrulewidth}{0pt}
 }

\makeFNbottom \makeatletter
\renewcommand\LARGE{\@setfontsize\LARGE{15pt}{17}}
\renewcommand\Large{\@setfontsize\Large{12pt}{14}}
\renewcommand\large{\@setfontsize\large{10pt}{12}}
\renewcommand\footnotesize{\@setfontsize\footnotesize{7pt}{10}}
\renewcommand\scriptsize{\@setfontsize\scriptsize{7pt}{7}}
\makeatother

\renewcommand{\thefootnote}{\fnsymbol{footnote}}
\renewcommand\footnoterule{\vspace*{1pt}%
\color{cream}\hrule width 3.5in height 0.4pt \color{black} \vspace*{5pt}}
\setcounter{secnumdepth}{5}

\makeatletter
\renewcommand\@biblabel[1]{#1}
\renewcommand\@makefntext[1]%
{\noindent\makebox[0pt][r]{\@thefnmark\,}#1}
\makeatother
\renewcommand{\figurename}{\small{Fig.}~}
\sectionfont{\sffamily\Large}
\subsectionfont{\normalsize}
\subsubsectionfont{\bf}
\setstretch{1.125} 
\setlength{\skip\footins}{0.8cm}
\setlength{\footnotesep}{0.25cm}
\setlength{\jot}{10pt}
\titlespacing*{\section}{0pt}{4pt}{4pt}
\titlespacing*{\subsection}{0pt}{15pt}{1pt}

\fancyfoot{}
\fancyfoot[LO,RE]{\vspace{-7.1pt}\includegraphics[height=9pt]{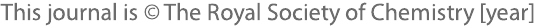}}
\fancyfoot[CO]{\vspace{-7.1pt}\hspace{13.2cm}\includegraphics{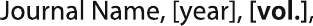}}
\fancyfoot[CE]{\vspace{-7.2pt}\hspace{-14.2cm}\includegraphics{head_foot/RF}}
\fancyfoot[RO]{\footnotesize{\sffamily{1--\pageref{LastPage} ~\textbar  \hspace{2pt}\thepage}}}
\fancyfoot[LE]{\footnotesize{\sffamily{\thepage~\textbar\hspace{3.45cm} 1--\pageref{LastPage}}}}
\fancyhead{}
\renewcommand{\headrulewidth}{0pt}
\renewcommand{\footrulewidth}{0pt}
\setlength{\arrayrulewidth}{1pt}
\setlength{\columnsep}{6.5mm}
\setlength\bibsep{1pt}

\makeatletter
\newlength{\figrulesep}
\setlength{\figrulesep}{0.5\textfloatsep}

\newcommand{\topfigrule}{\vspace*{-1pt}%
\noindent{\color{cream}\rule[-\figrulesep]{\columnwidth}{1.5pt}} }

\newcommand{\botfigrule}{\vspace*{-2pt}%
\noindent{\color{cream}\rule[\figrulesep]{\columnwidth}{1.5pt}} }

\newcommand{\dblfigrule}{\vspace*{-1pt}%
\noindent{\color{cream}\rule[-\figrulesep]{\textwidth}{1.5pt}} }

\makeatother

\twocolumn[
  \begin{@twocolumnfalse}
\vspace{3cm} \sffamily
\begin{tabular}{m{4.5cm} p{13.5cm} }

\includegraphics{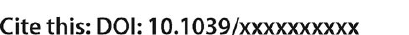} & \noindent\LARGE{Microwave-assisted coherent control of ultracold polar molecules with a ladder-type rotational states} \\
 & \vspace{0.3cm} \\
 & \noindent\large{Ting Gong,$^{\ast}$\textit{$^{a,b}$} Zhonghua Ji,$^{\ast}$\textit{$^{a,b}$}, Jiaqi Du,\textit{$^{a,b}$} Yanting Zhao,$^{\dag}$\textit{$^{a,b}$} Liantuan Xiao,\textit{$^{a,b}$} and Suotang Jia\textit{$^{a,b}$}} \\ & \vspace{-0.3cm} \\
\includegraphics{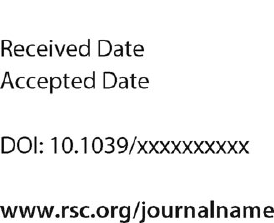} &\noindent\sffamily{We have demonstrated microwave-assisted coherent control of ultracold $^{85}$Rb$^{133}$Cs molecules with a ladder-type configuration of rotational states.
A probe microwave (MW) field is used to couple a lower state $X^1\Sigma^+(v=0, J=1)$ and a middle state $X^1\Sigma^+(v=0, J=2)$, while a control MW field couples the middle state and a upper state $X^1\Sigma^+(v=0, J=3)$. In the presence of the control field, the population of middle rotational states, $X^1\Sigma^+(v=0, J=2)$, can be reduced by a control MW field. Broadening of spectral splitting and shift of central frequency in this coherent spectrum are observed to be dependent on Rabi frequency of the control MW field. Applying Akaike's information criterion, we conclude that our observed coherent spectra happen through the crossover range of electromagnetically induced transparency and Aulter-Townes splitting as Rabi frequency of control field increases. Our work is a significant development in microwave-assisted quantum control of ultracold polar molecules with multilevel configuration, and also offers a great potential in quantum information based on ultracold molecules.}\\
&\vspace{0.3cm}
 \\
\end{tabular}

 \end{@twocolumnfalse} \vspace{0.6cm}

  ]

\renewcommand*\rmdefault{bch}\normalfont\upshape
\rmfamily
\section*{}
\vspace{-1cm}


\footnotetext{\textit{$^{a}$Shanxi University, State Key Laboratory of Quantum Optics and Quantum Optics Devices, Institute of Laser Spectroscopy, Wucheng Rd. 92,  030006 Taiyuan, China. }}
\footnotetext{\textit{$^{b}$Shanxi University, Collaborative Innovation Center of Extreme Optics, Wucheng Rd. 92, 030006 Taiyuan, China.}}
$^{\ast}$ These two authors contributed equally. \\
$^{\dag}$ zhaoyt@sxu.edu.cn

\rmfamily 

\section*{Introduction}
Coherent control of a three-level quantum system with external electromagnetic fields has promoted many intriguing quantum optical phenomena, including the widely-studied electromagnetically induced transparency (EIT) \cite{Harris1990,Boller1991,Fleischhauer2005} and Autler-Townes splitting (ATS) \cite{Autler1955} effects. Based on quantum interference characteristic,  EIT has witnessed numerous important applications in atomic system, including lasing without inversion \cite{Kocharovskaya1992}, high-precision magnetometry \cite{Staehler2001}, slow light propagation \cite{Kocharovskaya2001}, light storage \cite{Lukin2003} and quantum transistor \cite{Souza2013}. Meanwhile, besides of traditional spectral precision measurements including dipole transition moment \cite{Quesada1987} and lifetimes of coupled levels \cite{Garcia-Fernandez2005}, ATS has also some possible applications due to its coherence, such as recently proposed broadband quantum memory \cite{Saglamyurek2018}.

Along with these developments in atomic system, coherent control in three-level molecular systems also attracts researchers' attentions due to the existence of abundant rovibrational states in this system. Up to now, most experiments were implemented in homonuclear molecules, which are usually formed in heat pipe \cite{Qi1999a,Qi2002,Qi2006,Lazoudis2010,Lazoudis2011}. In this case, the existed Doppler broadening limits precise control of quantum states; large spontaneous rates of excited electronic states in such open system decrease pump-induced coherence; moreover homonuclear molecules have small transition dipole moments (TDM). Comparing with the homonuclear molecules, there are a few investigations on EIT or ATS effects of polar molecules in optical range with both theories \cite{Zhou2009,Singh2012a, Du2014josab, Ma2010jmo} and experiments in hollow-core photonic gap-fiber \cite{Ghosh2005} and in photonic microcell \cite{Light2009}. However, the disadvantages of Doppler broadening and large decay rate still exist. By comparison, ultracold polar molecules in the lowest vibronic state can overcome these two disadvantages: Doppler broadening can be ignored and spontaneous rate is largely suppressed. Thus this system provides an ideal platform to implement precise coherent control of multi-rotational states with frequency interval laying in microwave (MW) range. Comparing with optical range, MW transition has  higher accuracy and  better stability, supporting a higher resolution on quantum  control \cite{Aldegunde2009}. Besides of these advantages, MW coherent control of ultracold molecules with multi-rotational states also provides potential applications in quantum magnetism \cite{Gorshkov2011a}, topological phase \cite{Manmana2013} and synthetic dimension \cite{Sundar2018}.

Nowadays MW coherent control between adjacent two-level rotational states has been exhibited in different ultracold polar molecules formed by a variety of ways, including magnetoassociation followed by stimulated Raman adiabatic passage \cite{Ospelkaus2010,Will2016,Park2017,Gregory2016,Guo2018}, direct laser cooling in magneto-optical trap \cite{Williams2018} and short-range photoassociation (PA) recently demonstrated by our group \cite{Ji2020pccp}. These experiments provide a basis for expanding coherent control from two-level to three-level states.

In this work, we present coherent control of ultracold polar molecular in a ladder-type rotational system with probe and control MW fields.
Broadening of spectral splitting and shift of central frequency are observed to be dependent on the Rabi frequency of control MW field. The observed coherent spectra are found to happen through the crossover range of EIT and ATS by applying Akaike’s information criterion (AIC) \cite{Anisimov2011}.

\begin{figure}[t]
\centering
\includegraphics[width=0.45\textwidth]{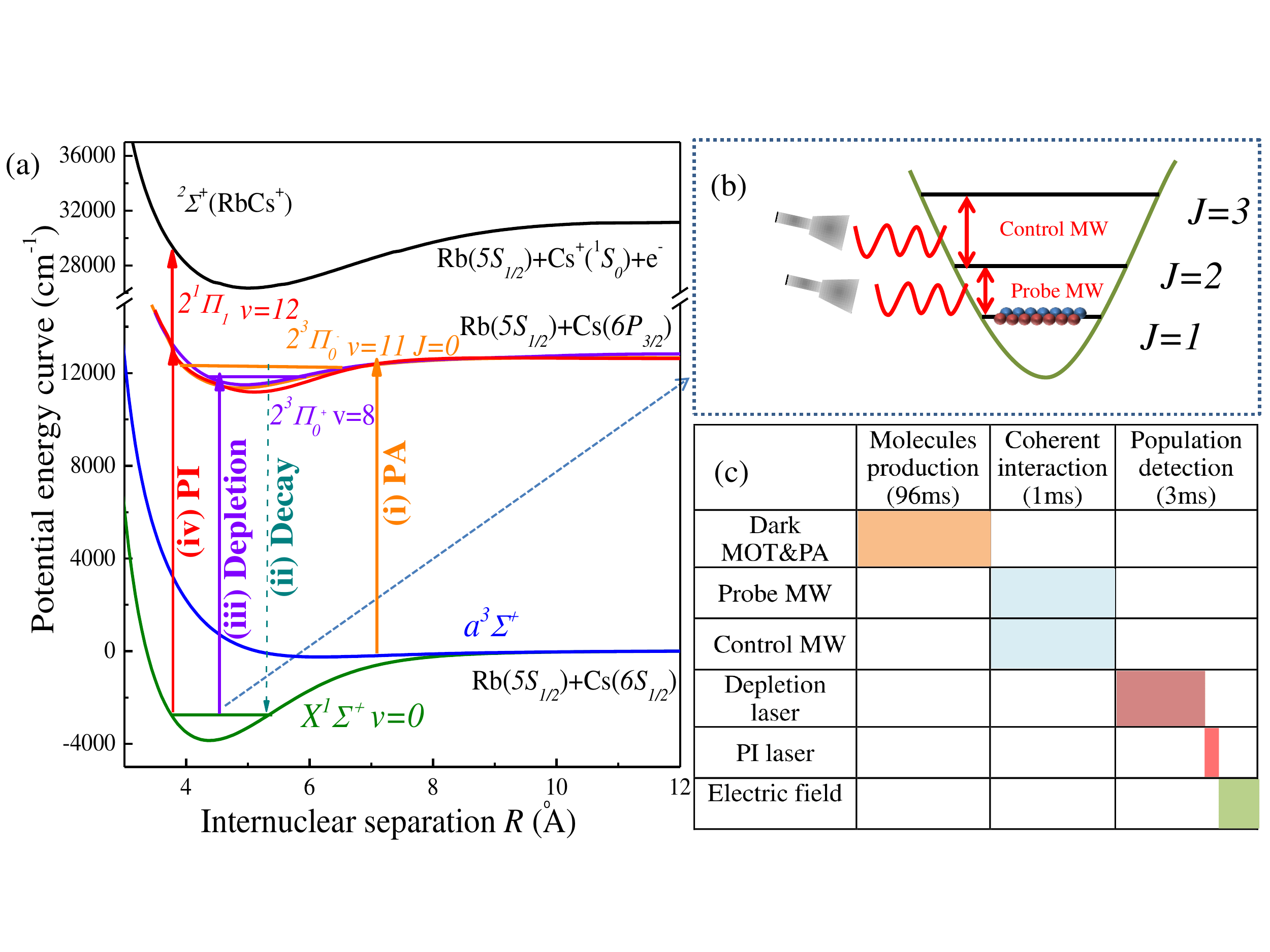}
\caption{(Color online) (a) Optical transitions of production and detection of $^{85}Rb^{133}Cs$ molecules. (b) A ladder-type rotational states for MW coherent control. (c) Time sequence. }
	\label{fig:seq}
\end{figure}

\section*{Experimental setup}

Figure~\ref{fig:seq} (a) shows the optical transitions we use to produce and detect $^{85}$Rb$^{133}$Cs molecules with rotational resolution. The intermediate level was chosen as $2^{3}\Pi_{0^{-}}(v = 11,J = 0)$ for that this state is theoretically expected to have only $J$ = 1 distribution for $v$ = 0 vibrational state \cite{Shimasaki2015} and has been verified in our measurement. Based on this simple distribution, two MW fields (labeled as probe and control MWs) were used to realize coherent control of rotational quantum states, as shown in Fig.\ \ref{fig:seq}(b). Two signal generators (Stanford Research Systems, SG386) and two power amplifiers (Mini-Circuits, ZHL-4W-422+) with two homemade antennas were used to provide probe and control MW fields. Two fast switches (ZFSWA2-63DR+) were used to switch on/off these MW pulses in less than 35ns.

The time sequence shown in Fig.\ \ref{fig:seq}(c) can be divided into three parts. The first  procedure is molecules production. $1\times10^{7}$ $^{85}$Rb atoms in 5$S_{1/2}$ ($F$ = 2) state have a density of $8\times10^{10}$ cm$^{-3}$ and $2\times10^{7}$ $^{133}$ Cs atoms in 6$S_{1/2}$ ($F$ = 3) state have a density of $1\times10^{11}$ cm$^{-3}$. A Ti: sapphire laser was simultaneously focused on these cold atomic samples to produce ultracold RbCs molecules in the $X^{1}\Sigma^{+}(v = 0,J = 1)$ ground state for 96 ms. The second procedure is coherent control of $^{85}$Rb$^{133}$Cs molecules with two MW fields. The third procedure is population detection of our interested rotational state by depletion spectroscopy technique \cite{Wang2007}. The details are described below. A photoionization (PI) laser is used to ionize  $^{85}$Rb$^{133}$Cs molecules in the $X^{1}\Sigma^{+}(v = 0)$ vibrational state and an electric field accelerates these photoionized ions to a pair of microchannel plates. The repetition rate of total experiment sequence is chosen to be  10 Hz, accordant with the period of PI laser pulse.
We fix the wavelength of PI laser at 651.83 nm to photoionize the formed $X^{1}\Sigma^{+}$($v$ = 0) state molecules via $2^{1}\Pi_{1}$($v$ = 12) intermediate state. Here the depletion laser frequency is locked at 466.8575 THz, which is resonant with transitions from $X^{1}\Sigma^{+}(v = 0,J = 2$) to $2^{3}\Pi_{0^{+}}(v = 8,J = 1$) \cite{Yuan2015} to measure the population of $J=$ 2 state. Once the depletion laser is on the resonant transition, the intensity of detected molecular ion from $v = 0$ vibrational state will be depleted. The fractional depletion can indicate the rotational population.

\section*{Experimental results and analysis}
\begin{figure}[t]
\centering
\includegraphics[width=0.45\textwidth]{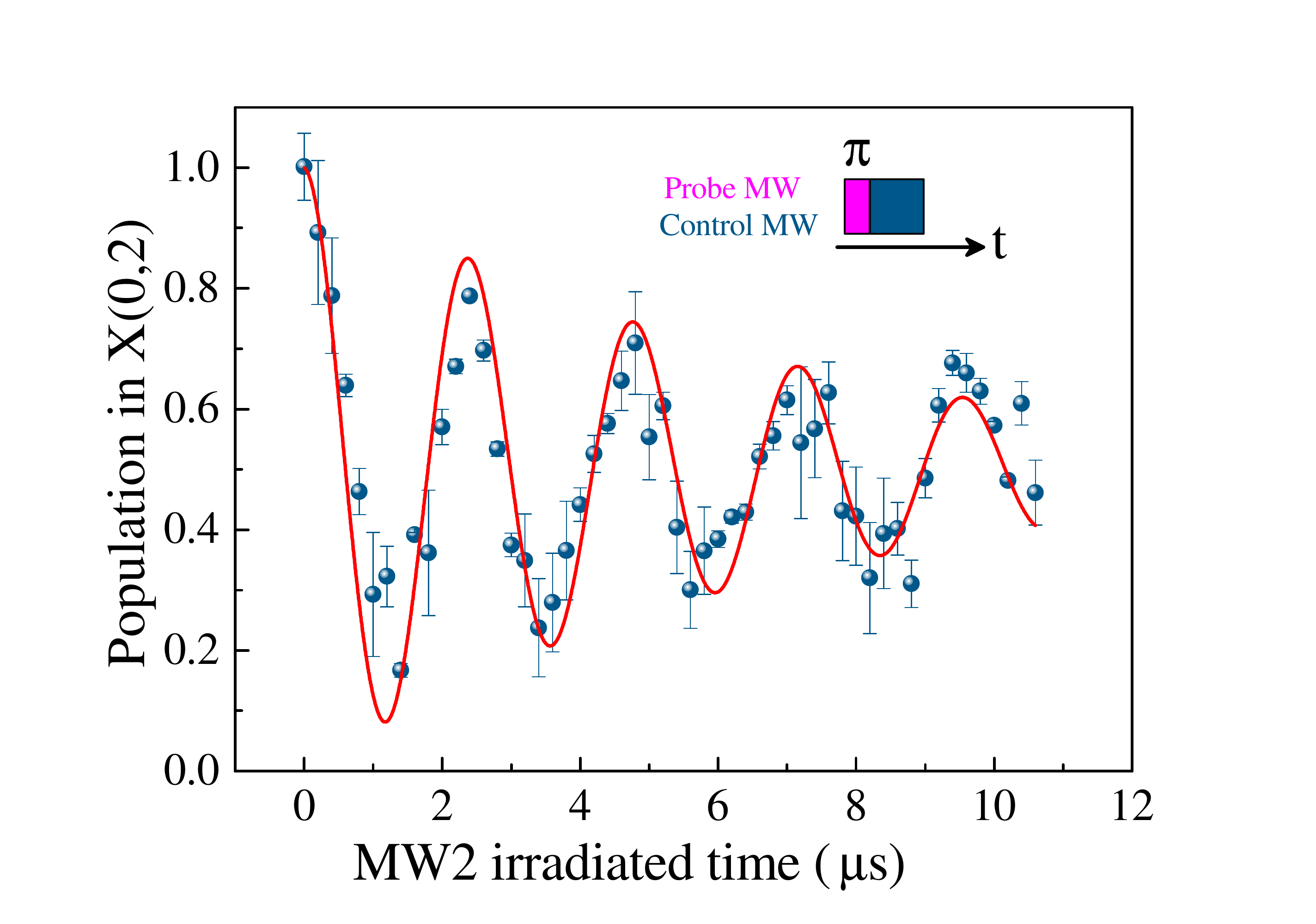}
\caption{(Color online) Rabi oscillation between $X(v=0,J=2)$ and $X(v=0,J=3)$ states under a control MW  field. Ultracold $^{85}$Rb$^{133}$Cs molecules are initially populated in the $J=1$  state by spontaneously decay, then transferred to $J=2$ by a $\pi$ pulse of probe MW, followed by irradiation of control MW.}
	\label{fig:Rabi}
\end{figure}

To implement coherent control on a three-level system, we need to know  Rabi frequencies coupled by probe and control fields and the corresponding coherence decay rates. Rabi  frequency coupled by probe field $\Omega_{p}$ and coherence decay rate between $J$ = 1 and $J$ = 2 states $\gamma_{12}$ have been derived from the Rabi oscillation between them using one MW field \cite{Ji2020pccp}. To observe Rabi oscillation between $J$ = 2 and $J$ = 3 states, here two MW fields are implemented. We firstly irradiate molecules in $J$ = 1 state with a $\pi$ pulse of probe MW aiming at transferring molecules to $J$ = 2 state. Then we turn on only control MW with a varied irradiation time, followed  by recording the population of molecules in $J$ = 2 state.
Figure \ref{fig:Rabi} shows our observed Rabi oscillation between $J$ = 2 and $J$ = 3 rotational states. In this measurement, the frequency of control MW is resonant with the transition between $J$ = 2 and $J$ = 3 states (2982.82 MHz) \cite{Li2018}. We fit these data with  a simple trigonometric function with an exponential decay

\begin{eqnarray}
N=Acos(\Omega_{c}t)exp(-\gamma_{23}t)+C
\label{eq1}
\end{eqnarray}

Here $A$ and $C$ are oscillation amplitude and offset, with fitted values of 0.50(4) and 0.50(1) respectively.  Rabi frequency of pump MW field $\Omega_{c}$ and coherence decay rate between $J=$ 2 and $J=$ 3 states $\gamma_{23}$ are fitted to be 0.42(1) MHz and 0.15(3) MHz respectively.

The measured value of $\Omega_{c}$ allows us to derive TDM between $J=$ 2 and 3 states by $\mu_{23}$=$\sqrt{(cn\varepsilon_{0}\pi r^2 \hbar^2)/2P}\cdot\Omega_{c}$ = 0.57(1) Debye, where $r$ is detector radius of MW power meter (1.15 cm) and $P$ is irradiated MW power at molecular sample, which is measured by a microwave power meter with 0.3 mW. To compare with a calculated value, we utilize the same operation as dealing with TDM between $J=$1 and 2 states $\mu_{12}$ in Ref. \cite{Ji2020pccp}, but here for $J$ = 2 and 3 states $\mu_{23}$ is expected to be $\mu_{23}=\sqrt{(9/35)}$$\mu_{0}$ \cite{Brink-Satchler1994Book}, where $\mu_{0}$ is adopted to be 1.225(11) Debye \cite{Gregory2016} by  simplifying it to be the value between $|J=2,M=0>$ and $|J=3,M=0>$ under a plane-polarized light with the electric vector along a single direction \cite{Hougen1970Book} and by considering that $\mu_{0}$ for $^{85}$Rb$^{133}$Cs is only about 9 parts in 10$^{7}$ smaller than for $^{87}$Rb$^{133}$Cs\cite{Ji2020pccp}. Thus our measured value is consistent with this calculated value in view of the uncertainty in the measured MW power.

It always takes lots of time to acquire a graph such as Fig.\ \ref{fig:Rabi}, due to low signal-to-noise ratio (SNR) when MW power is too weak. Thus we use the relationship $\Omega\propto\sqrt{P}$ to obtain an expected Rabi frequency by changing MW power. To obviously observe coherent effect of a three-level system, $\Omega_{p}$ is expected to be  smaller than $\gamma_{12}$ \cite{Anisimov2011,Abi-Salloum2010a}, which has been measured to be 0.67 MHz \cite{Ji2020pccp}. Hence in the following context, we set $\Omega_{p}$ only 17 kHz where an obvious SNR can still be obtained.

\begin{figure}[t]
\centering
\includegraphics[width=0.45\textwidth]{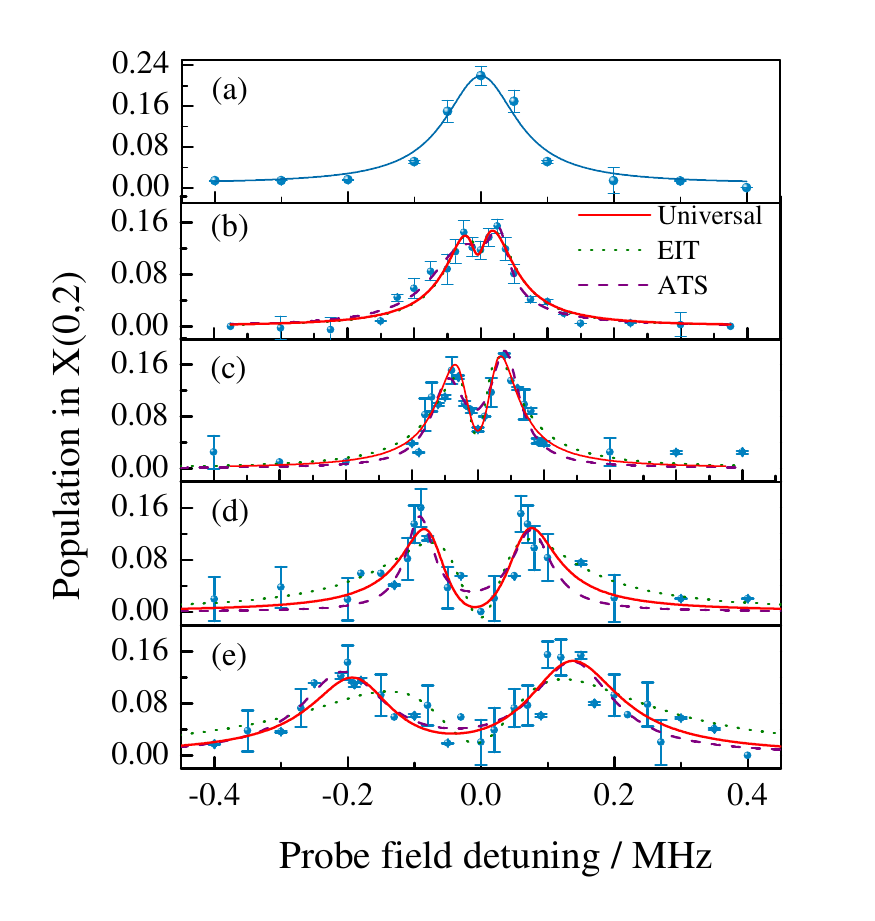}
\caption{(Color online) Coherent spectra of ultracold $^{85}$Rb$^{133}$Cs molecules with ladder-type rotational states.
The population of J=2 is measured in the absence (a) and presence (b-e) of control field with the detuning of the probe field. The curve in (a) is a Lorentz fitting, red solids in (b-e) spectra are the fitting results based on the imaginary part of Eq.\ \ref{eq2} under approximation
of $\Omega_{ce}\sim\Omega_{c}$. The green line and purple line are the fitting lines using EIT model and AT model.
The Rabi frequency of control field are 0, 0.055, 0.09, 0.26, 0.42 MHz for (a-e) respectively.}
	\label{fig:EIT}
\end{figure}

Figure \ref{fig:EIT} plots $J$ = 2 rotational population excited by the probe field under different Rabi frequencies of control MW field. In the absence of the control MW displayed in Fig.\ \ref{fig:EIT}(a), the spectrum shows an expected typical Lorentz lineshape as a function of probe field detuning $\delta_{p}$. The fitted central frequency is 1988.604 (4) MHz, which is resonant transition of $J$ = 1 and $J$ = 2 states. We set this value to be the zero point for all the horizontal coordinates in Fig.\ \ref{fig:EIT}. In the presence of control MW shown in Fig.\ \ref{fig:EIT}(b-e), the population of $J$ = 2 is reduced when control MW power increases. As the population reduction of $J$ = 2 state stands for the absorption decreasing of probe field, Fig. \ref{fig:EIT}(b-e) exhibits enhancing transparency of the probe MW field induced by the control MW field.

In 1995 Gea-Banacloch $et$ $al.$ \cite{Gea-Banacloche1995} theoretically studied laser absorption in atomic system with both ladder-type and $\Lambda$-type. But for polar molecular system there may need different theory due to the existence of  permanent dipole moment (PDM). In 2009 Zhou $et$ $al.$ \cite{Zhou2009} firstly theoretically studied EIT in a $\Lambda$-type polar molecule system.
They found that multi-photon transition case with virtual level(s) can happen in this kind of system, and an oscillatory feature will appear in the common (1+1)-transition processes if the PDMs of molecular states coupled by control field are comparable. Combining the obtained susceptibility in Eq. 22 of Ref. \cite{Zhou2009} and density-matrix element in Eq. 3 of Ref. \cite{Gea-Banacloche1995}, we write the susceptibility for a ladder-type polar molecule with (1+1)- transition as

\begin{eqnarray}
\chi=\frac{K(\delta_{p}+\delta_{c}+i\gamma_{13})}
{(\gamma_{12}-i\delta_{p})(\gamma_{13}-i\delta_{p}-i\delta_{c})+\Omega_{ce}^{2}/4}.
\label{eq2}
\end{eqnarray}

In this universal expression, we define $\Omega_{ce}=2\Omega_{c}J_{1}(Z_{32}^{c})/Z_{32}^{c}$ as effective Rabi frequency related to control field, in which $\Omega_{c}=\mu_{23}E_c/\hbar$  is Rabi frequency and $J_{1}(Z_{32}^{c}$) is the first Bessel function of integer order 1 on the variable $Z_{32}^{c}=(\mu_{3}-\mu_{2})E_{c}/\omega_{c}$. Here $\hbar$ is the reduced Planck constant, $\mu_{3}$ and $\mu_{2}$ are respectively PDMs of states coupled by the control field, while $\mu_{23}$ is their TDM.  $E_{c}$ is amplitude of coupling field. As mentioned before, if $\mu_{3}$ is  comparable with $\mu_{2}$, the coherent phenomenon will be complicated even for (1+1)-transition because of an oscillatory feature in Bessel function as a function of $E_{c}$ \cite{Zhou2009}. Fortunately, here the difference between $J=$3 and 2 states in $v=$0 vibronic state is only 6$D_{\mu0}=1.4\times$10$^{-6}$ Debye \cite{Ji2020pccp}, resulting that $\Omega_{ce}$ equals to $\Omega_{c}$ under experimentally available MW power.


It is known that the imaginary part of susceptibility  leads to absorption characteristic of probe field, corresponding to molecular population variation in the states coupled by probe field. By setting $\gamma_{12}$, $\gamma_{13}$, $\Omega_{c}$, $\delta_{c}$ and $K$ as variables, the fitting results are shown with red solids in Fig.\ \ref{fig:EIT}(b-e). $K$ is only a coefficient influencing spectral amplitude. For all the graphs, $\gamma_{12}$ and $\gamma_{13}$ are expected and truly verified to be unchanged, with averaged values of 0.2(1) MHz and 0.02(1) MHz respectively.

\begin{figure}[t]
\centering
\includegraphics[width=0.45\textwidth]{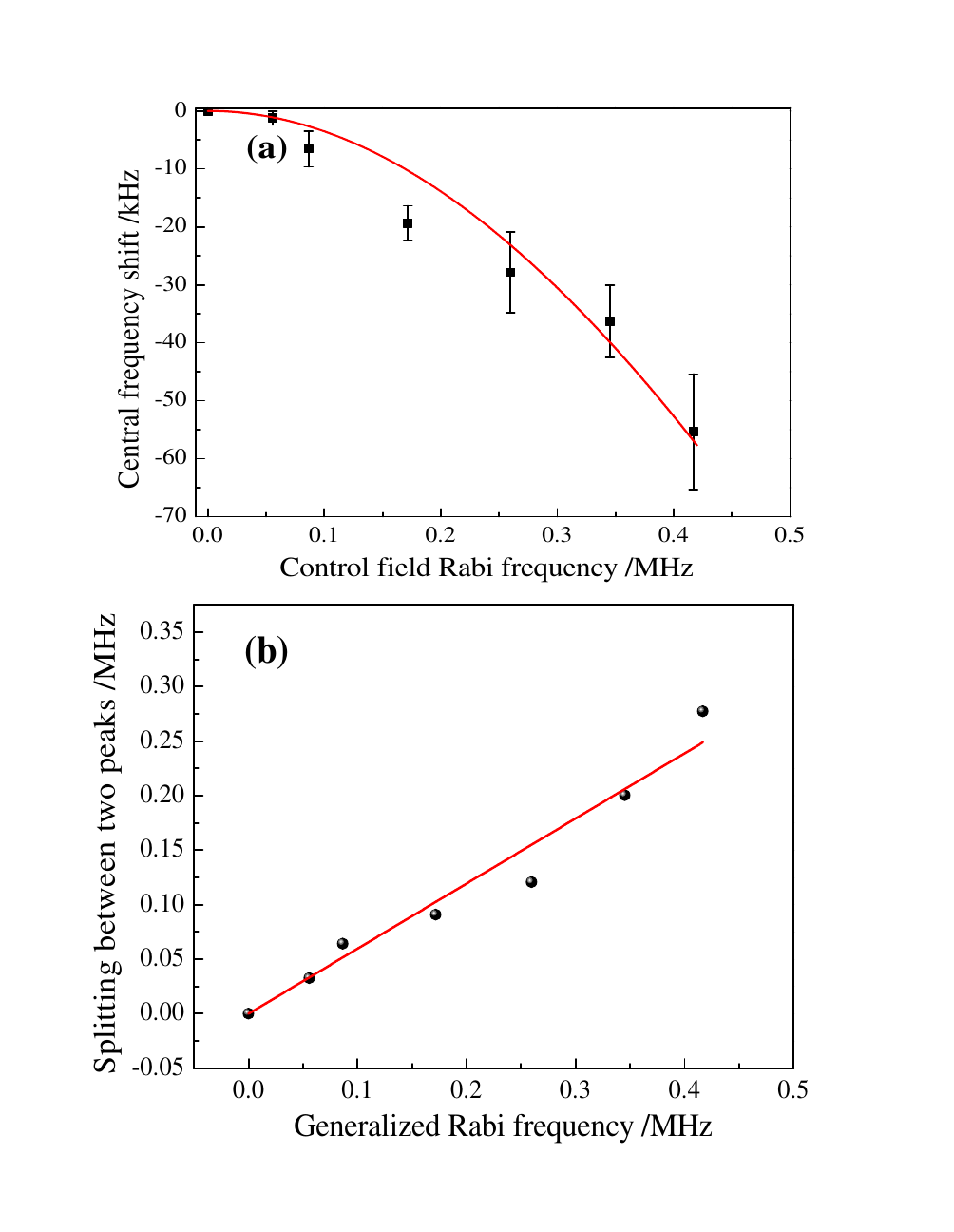}
\caption{(Color online) Central frequency shift (a) in the coherent effect and control MW field-induced spectrum splitting (b). The curve in (a) is fitting result based on Eq.\ \ref{eqn3}, the lines in (b) is linear fitting.}
	\label{fig:split-shift}
\end{figure}

Regards to the detuning of control MW $\delta_{c}$, a frequency shift induced by Rabi frequency of control MW field is observed, shown in Fig.\ \ref{fig:split-shift}(a).
We attribute it to the  A.C. Stark effect or light shift. As $J$ = 2 state is influenced mainly by the strong control MW field (the influence of weak probe field can be ignored), its energy shift induced by $\Omega_{c}$ can be written below from an ideal two-level system \cite{Foot}:
\begin{eqnarray}
\delta_c=\frac{\delta_{c0}}{2}-\frac{\sqrt{\delta_{c0}^{2}+\Omega_{c}^{2}}}{2}. \label{eqn3}
\end{eqnarray}

Here the initial detuning of control MW $\delta_{c0}$ is fitted to be 710(60) kHz. The factor influencing this detuning  arises mainly from unresolved hyperfine structure. In fact the molecules in  rotational state $J$ have $(2J+1)(2I_{Rb}+1)(2I_{Cs}+1)$ hyperfine sublevels, where nuclear angular numbers $I_{Rb}$ = 5/2 for $^{85}$Rb and $I_{Cs}$ = 7/2 for $^{133}$Cs. According to the Ref. \cite{Aldegunde2017PRA}, the dominant hyperfine interactions in states with $J>0$ are nuclear electric quadrupole interactions. The allowed microwave lines for $J=2\rightarrow3$ in the absence of control field are expected to be spread over about 567 kHz \cite{JM}, which is comparable  with $\delta_{c0}$.

Except for the central frequency shift, the spectrum splitting in Fig.\ \ref{fig:EIT}(b-e) is observed to increase with the Rabi frequency of control field. This phenomenon is similar as many investigations \cite{Kim2020, Zhang2013} with the relationship $\Delta_{sp} = \alpha\sqrt{\Omega_{c}^{2}+\delta_{c}^{2}} = \alpha\Omega_{c}$$'$. Here $\Omega_{c}$$'$ is generalized Rabi frequency of control MW field and $\alpha$ is a coefficient. We plot the generalized Rabi frequency-dependent of the observed spectrum splitting in Fig.\ \ref{fig:split-shift}(b) and the $\alpha$ is extracted to be 0.60(3).

So far, we have used an universal formula to analyze our observed  coherence spectra and do not assign these spectra into EIT or ATS, which are two extremes of three-level coherent spectroscopy. Recently, investigations on discernment between EIT and ATS have become an active topic \cite{Anisimov2011, Abi-Salloum2010a, Giner2013, Zhu2013,Liu2016,Peng2014,He2015,Liu2016a}. Among these works,  AIC is proposed to discriminate EIT and ATS from an experimental viewpoint \cite{Anisimov2011,Giner2013,Liu2016,Peng2014,He2015,Liu2016a}. It has been employed widely to quantitatively determine the relative weights of the effects of EIT and ATS in various systems, including cold atoms \cite{Giner2013}, mechanical oscillators \cite{Liu2016}, whispering gallery mode microcavities \cite{Peng2014}, plasmonic waveguide-resonators \cite{He2015} and superconducting quantum circuits \cite{Liu2016a}. Here we apply AIC  to ultracold polar molecule.

The lineshapes of EIT and ATS can be expressed as $A_{EIT}=C_{+}/(1+(\delta-\epsilon)^{2}/(\gamma_{+}^{2}/4))-C_{-}/(1+\delta^{2}/(\gamma_{-}^{2}/4))$ and $A_{ATS}=C_{1}/(1+(\delta+\delta_{1})^{2}/(\gamma_{1}^{2}/4))+C_{2}/(1+(\delta-\delta_{2})^{2}/(\gamma_{2}^{2}/4))$ respectively \cite{Giner2013}. Here $C_{1}$, $C_{2}$ and $C$$_{\pm}$ stand for the amplitudes of Lorentzian curves,  $\gamma_{1}$, $\gamma_{2}$ and  $\gamma_{\pm}$ for the linewidths respectively, $\epsilon$, $\delta_{1}$ and $\delta_{2}$ for the detunings from resonant frequency. The fitting results are shown with green and purple curves for EIT and ATS models in Fig.\ \ref{fig:EIT}(b-e). The per-point AIC contribution of EIT (ATS) model is quantitated by $\omega_{EIT(ATS)}=exp(-I_{EIT(ATS)}/2N)/[exp(-I_{EIT}/2N)+exp(-I_{ATS}/2N)]$, where \emph{N} is the data number of each spectrum and $I=N\cdot log(\sum_{j=1}^{N}\epsilon_{j}^{2}/N)+2M$ . Here \emph{M} is the number of fitting variables and $\epsilon_{j}^{2}$  is the residual of experimental measurement from the fitted model. The weights of EIT model in Fig.\ \ref{fig:EIT}(b-e) are 0.57, 0.50, 0.48, 0.45 while the weights of ATS are the values subtracted by one as $\Omega_{c}$ increases. It means that EIT dominates the lineshape of the spectrum when $\Omega_{c}$ is lower, while ATS dominates when $\Omega_{c}$ is larger.

\section*{Conclusion}

In conclusion, we have reported microwave-assisted experimental realization of coherent spectroscopy of ultracold polar molecules with a ladder-type rotational states. The coherent control is performed by monitoring the population of middle rotational state in the presence of control MW field. The spectrum splitting is proportional to the Rabi frequency of control field. The observed frequency detuning in coherent spectra is attributed to the A.C. Stark effect induced by the control MW field. The initial detuning mainly arises from unresolved hyperfine structure. By employing AIC to the experimental spectra, it was found that our observed coherent spectra happen through the crossover range of EIT and ATS. To observe ATS-dominated spectrum, it is expected to increase irradiated power of the control field. To observe EIT-dominated spectrum, it is expected to decrease probe field power and ensure enough SNR by improving detection sensitivity and increasing molecule density. Our presented quantum interference phenomenon of ultracold polar molecules is meaningful for microwave-assisted coherent control with multilevel configuration.

\section*{Acknowledgments}
We acknowledge Jeremy M. Hutson for theoretical assistants, Jinze Wu for useful discussions. This work was supported by National Key R$\&$D Program of China (Grant No. 2017YFA0304203), Natural
Science Foundation of China (Nos. 61675120, 61875110, 12074231, 12034012), NSFC Project for Excellent Research Team
(No. 61121064), Shanxi $``$1331 Project$"$ Key Subjects Construction, PCSIRT (No. IRT$_{-}$17R70),
111 project (Grant No. D18001).


\scriptsize{
\bibliography{EIT} 
\bibliographystyle{rsc} } 

\end{document}